\documentclass[useAMS,usenatbib,usegraphicx,letterpaper]{mn2e}

\newcommand{\etal}{et al.}

\title[The {\it XMM--Newton} view of 3C~109] {Fe emission and ionized excess
  absorption in the luminous quasar 3C~109 with XMM--Newton}
\author[G.\ Miniutti \etal]
{G.~Miniutti$^1$\thanks{miniutti@ast.cam.ac.uk}, D.~R.~Ballantyne$^2$,
  S.~W.~Allen$^3$, A.~C.~Fabian$^1$ and R.~R.~Ross$^4$\\
  $^1$ Institute of Astronomy, Madingley Road, Cambridge CB3 0HA \\
  $^2$ Department of Physics, University of Arizona, 1118 East 4th
  Street, Tucson, AZ 85721 USA\\
  $^3$ Kavli Institute for Particle Astrophysics and Cosmology,
  Stanford University, 382 Via Pueblo Mall, Stanford, CA 94305--4060, USA\\
  $^4$ Physics Department, College of the Holy Cross, Worcester, MA
  01610, USA}

\pagerange{\pageref{firstpage}--\pageref{lastpage}}
\pubyear{2001}

\usepackage{times}

\begin{document}

\label{firstpage}

 \maketitle

\begin{abstract}
  We report results from an {\it XMM--Newton} observation of the
  broad--line radio galaxy 3C~109 (z=0.3056).  Previous {\it ASCA}
  data revealed the presence of a broad iron line from the accretion
  disc with which the {\it XMM--Newton} spectrum is fully consistent.
  However, although improving the {\it ASCA} constraints on the line
  parameters, the quality of the data is not high enough to
  distinguish between an untruncated accretion disc extending down to
  small radii close to the black hole and a scenario in which the
  innermost 20--30 gravitational radii are missing. For this reason,
  our results are model--dependent and the hard data can be modelled
  equally well by considering an absorption scenario in which a large
  column of neutral gas partially covers the X--ray continuum source.
  However, the absorber would have to comprise hundreds/thousands very
  compact clouds close to the X--ray source, which seems rather
  extreme a requirement. The 2--10~keV intrinsic luminosity of 3C~109
  is of the order of 2--3$\times 10^{45}$erg~s$^{-1}$ regardless of
  the adopted model. A recent black hole mass estimate of $\sim 2
  \times 10^8~M_\odot$ implies that $L_{\rm{bol}} /L_{\rm{Edd}} > 1$.
  If partial covering is excluded, the observed reflection fraction
  (of the order of unity), steep photon index (1.86), and Fe line
  equivalent width (about 100~eV) all suggest to exclude that the
  X--ray continuum is strongly beamed indicating that the large
  Eddington ratio is associated with a radiatively efficient accretion
  process and making it unlikely that the innermost accretion disc is
  replaced by a thick radiatively inefficient medium such as in
  advection--dominated accretion models. We also confirm previous
  findings on the detection of low energy absorption in excess of the
  Galactic value, where we find excellent agreement with previous
  results obtained in X--rays and at other wavelengths (optical and
  infrared).  The better quality of the {\it XMM--Newton} data enables
  us to attribute the excess absorption to slightly ionized gas in the
  line of sight, located at the redshift of 3C~109.  The most likely
  interpretation for the excess absorption is that the line--of--sight
  is grazing the obscuring torus of unified models, which is
  consistent with the inclination inferred from the Fe line profile
  (about 40$^\circ$) and with the hybrid radio--galaxy/quasar nature
  of 3C~109.
\end{abstract}


\section{Introduction}

According to orientation--dependent unified models, radio--loud
quasars and radio galaxies belong to the same parent population and
their observed properties depend on our viewing angle with respect to
the obscuring torus and jet axis (e.g. Orr \& Browne 1982; Barthel
1989; Antonucci 1993). The dusty torus hides the active nucleus and
broad--line region in radio galaxies, whereas these are viewed
directly in quasars. The half opening angle of the torus is thus the
critical angle (about $45^\circ$, e.g. Barthel 1989) at which the
transition between radio galaxies and quasars is thought to occur.
However, given the observed increase of quasar fraction with distance
and luminosity (Singal 1996; Willot et al. 2000), the opening angle is
likely to increase with luminosity (e.g. the receding torus model,
Lawrence 1991; Simpson 2003 for a review) allowing the broad--line
region to be seen at the highest luminosities (but see e.g. Haas et
al. 2004).

3C~109 is an example of a luminous active galactic nucleus (AGN) with
observed properties placing it at the boundaries between broad--line
radio galaxies (BLRGs) and quasars. In the optical 3C~109 appears as a faint
galaxy at z=0.3056 (Spinrad et al. 1985) with an unresolved nucleus.
Early optical studies revealed a large Balmer decrement and a
red continuum (Yee \& Oke 1978; Grandi \& Osterbrock 1978). Rudy et
al.  (1984) found strong and non--variable optical polarisation (about
8 per cent) in both the optical continuum and H$\alpha$ emission. Such
a level of polarisation, if due to transmission through aligned dust
grains, is consistent with a large intrinsic reddening of the order of
E(B-V)=1.0--1.5, a result later consolidated through
spectro--polarimetric observations by Goodrich \& Cohen (1992) who
derived a total continuum reddening of E(B-V)~$\simeq$~1.2.
Subsequent spectro--photometric observations in the infrared by Rudy,
Puetter \& Mazuk (1999) revealed an in situ reddening of E(B-V)
$=0.77\pm 0.09$ to be added to the Galactic one of E(B-V) $=0.57$
estimated through the COBE/IRAS infrared maps\footnote{As pointed out
  by Rudy, Puetter \& Mazuk (1999) the COBE/IRAS estimate is much
  larger that the value E(B-V) $=0.27\pm 0.03$ derived from the H I
  measurements of Burstein \& Heiles (1982) but should be more
  accurate since it is based on direct observation of dust.} of
Schlegel, Finkbeiner \& Davis (1998).

3C~109 is also a bright source in the X--ray band with a 2--10~keV
intrinsic luminosity of 2--5 $\times 10^{45}$erg~s$^{-1}$ (Allen \&
Fabian 1992; Allen et al 1997) making it one of the most X--ray
luminous objects at $z<0.5$. Absorption is seen in the X--rays in
excess of that expected from Galactic material, confirming the optical
and infrared studies. {\it ROSAT PSPC} (Allen \& Fabian 1992) and {\it
  ASCA} (Allen et al 1997) data were used to infer a {\emph{total}}
X--ray column density of N$_H \sim 7\times 10^{21}$~cm$^{-2}$, in good
agreement with the optical reddening (N$_H = E(B-V)\cdot 5.8\times
10^{21} \simeq 7.7\times 10^{21}$; Bohlin, Savage \& Drage 1978). The
{\it ASCA} data also revealed the presence of a broad iron (Fe)
K$\alpha$ emission line (Allen et al 1997). The line was consistent
with being produced in the accretion disc but the emitting region was
only poorly constrained with an upper limit on the inner disc radius
of $\sim$140 gravitational radii. When interpreted as due to
reflection from the accretion disc, the broad Fe line profile
suggested an observer inclination $\theta > 35^\circ$ (for a neutral
6.4~keV Fe line).

At radio wavelengths, 3C~109 is classified as a FR~II lobe--dominated
radio galaxy with steep spectrum (Laing, Riley \& Longair 1983) and
large luminosity ($\sim 9\times 10^{26}$~W~Hz$^{-1}$; Bennet et al.
1986). The large--scale radio structure of 3C~109 consists of two
symmetric lobes with hot spots and core emission. A faint and clearly
one--sided jet is present at the parsec scale (Giovannini et al 1994).
By measuring the jet/core flux ratio in VLA data Giovannini et al. (1994)
could derive a value of $\theta < 34^\circ$ for the angle between the
radio axis and the line of sight.  VLBI observations of the
jet/counter--jet ratio however indicate a less tight constraint of
$\theta < 56^\circ$.  By combining the radio constraints on the
orientation with that from the broad Fe K line ($\theta > 35^\circ$),
3C~109 seems to lie at the high--inclination end of the quasar class,
which could also explain the observed excess absorption in terms of a
line of sight passing through the edge of the obscuring torus.

The {\it ASCA} detection of the broad Fe line in 3C~109 represents the
exception rather than the rule for radio--loud objects (e.g.
Eracleous \& Halpern 1998; Wo\'{z}niak et al. 1998; {Eracleous,
  Sambruna \& Mushotzky 2000; Grandi et al. 2006). The weak disc
  reflection signatures in radio--loud sources possibly indicate that
  the X--ray continuum is beamed away from the reflector (e.g.
  Beloborodov 1999) or the presence of a truncated accretion disc
  geometry in which the inner regions of the standard radiatively
  efficient and geometrically thin disc are replaced by a thick and
  radiatively inefficient medium such as in an advection--dominated
  accretion flow (Narayan \& Li 1995; see also Meier 1999; 2001 for
  the connection with jets). An alternative explanation is that the an
  untruncated disc is present, but highly ionized, thus producing
  weaker reflection signatures than expected in the neutral case
  (Ballantyne et al. 2002). As mentioned, the {\it ASCA} result on
  3C~109 is consistent with both an untruncated disc and one in which
  the innermost 150 gravitational radii are missing thereby not
  providing a clear picture of the innermost accretion flow, while the
  relatively steep X--ray slope ($\Gamma \simeq 1.86$) was used to
  exclude a strong contribution from a beamed jet component. So far,
  {\it XMM--Newton} observations of BLRGs have produced inconclusive
  results (e.g. Gliozzi et al. 2004; Ballantyne et al.  2004; Lewis et
  al.  2005) with some exceptions such as e.g. the BLRG 4C~74.26 and
  the radio--loud quasar PG~1425+267 in which evidence for an
  untruncated disc has been recently reported by Ballantyne \& Fabian
  (2005) and Miniutti \& Fabian (2006), respectively.

Here we present results from a new $\sim$~40~ks X--ray observation of
3C~109 with {\it XMM--Newton}. The higher sensitivity of the {\it
  XMM--Newton} CCD detectors with respect to any other X--ray mission
allows us to study the Fe band with unprecedented detail in search for
a confirmation of the {\it ASCA} detection of a broad Fe emission
line. We also present results on the excess absorption seen at low
energies with the aim of providing a self--consistent picture of the
environment of 3C~109 in the X--ray, optical, and infrared bands.

\section{The XMM--Newton observation}

3C~109 was observed by {\it XMM--Newton} on 2005 February 3--4 during
revolution 944. Source/background light curves and spectra for the
EPIC detectors were extracted from circular regions centred on the
source and close to it respectively. High background is present in the
last few ks of the observation which are excluded from the analysis.
Another period of relatively high background is instead included after
verifying that it does not affect the spectra in any noticeable way.
After filtering, the net exposure is $\sim$34~ks for the pn and
$\sim$37~ks for both MOS cameras. The pn spectrum was grouped as to
have a minimum of 50 counts per bin, while the MOS spectra were
grouped to have 25 counts per bin. 

It should be stressed that X--ray emission from the radio lobes has
been detected with {\it Chandra} (Croston et al 2005). The {\it
  XMM--Newton} spatial resolution does not allow us to distinguish
between core and lobe X--ray emission.  However, Croston et al. (2005)
could measure an upper limit of $3.7$~nJy for the 1~keV flux density
due to the lobes in 3C~109, while the 1~keV flux density of the core is
$896^{+129}_{-109}$~nJy (Hardcastle, Evans \& Croston 2006). Given
that the X--ray core is slightly brighter during the present {\it
  XMM--newton} observation than in the previous {\it Chandra} one (see
Fig.~6 below), the lobe contribution to the {\it XMM--Newton} spectrum
is likely limited to less than 0.3--0.4 per cent, making the spectrum
analysed below completely dominated by the core emission.

\begin{figure}
\begin{center}
 \includegraphics[width=0.32\textwidth,height=0.44\textwidth,angle=-90]{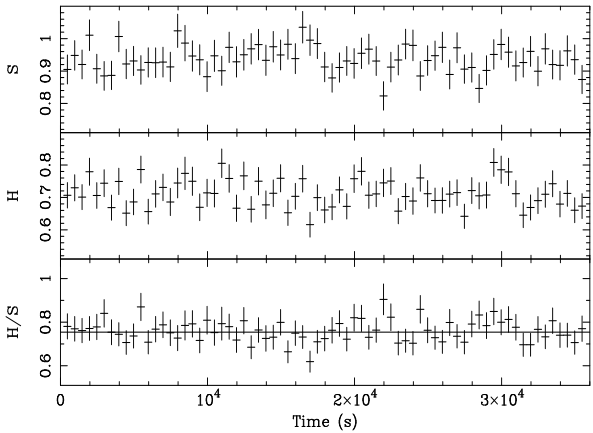}
\end{center}
\caption{We show the background--subtracted EPIC pn light curves in a
  soft (S: 0.2--2~keV) and a hard (H: 2--10~keV) band and the
  resulting hardness ratio (H/S). Each time--bin is 500~s long. Both
  light curves exhibit very small amplitude variations (at the few per
  cent level) but a fit with a constant is acceptable for both the S
  and the H bands. In the bottom panel, we also show the best--fit
  constant model for the harness ratio (H/S=$0.754\pm 0.006$) which
  results in a $\chi^2=53.5$ for 70 degrees of freedom.}
\end{figure}

In Fig.~1, we show the EPIC pn light curve in a soft (S: 0.2--2~keV,
top panel) and in a hard (H: 2--10~keV, middle panel) energy band
together with the resulting hardness ratio (H/S, bottom panel). The
two light curves are consistent with being constant over the
XMM--Newton exposure. A constant fit to the hardness ratio is also
totally acceptable ($\chi^2=53.5$ for 70 degrees of freedom) excluding
any strong spectral variation during the observation. Since 3C~109 does
not exhibit any sign of short--timescale flux or spectral variability,
we concentrate on the time--averaged X--ray spectrum of the source.

\section{The 2--10~keV spectrum and Fe K band}

We start our analysis by considering the hard spectrum of 3C~109 in
the 2--10~keV band (observed frame, i.e. 2.6--13.06~keV in the
rest--frame, z=0.3056). We consider the EPIC--pn data only because of the higher
sensitivity of the pn camera with respect to MOS in the Fe band. A
simple absorbed power law model provides a good description of the
hard spectrum for a power law photon index of
$\Gamma=1.67^{+0.07}_{-0.05}$. Neutral photoelectric absorption is
also seen even above 2~keV and we measure a column density of
$N_H=3.5^{+2.5}_{-1.7}\times 10^{21}$~cm$^{-2}$. The actual value of
the column density will be better constrained by including the soft
data, and we defer any discussion to a further analysis below.
The overall fit is acceptable with $\chi^2=367$ for 398 degrees of
freedom (dof), but visual inspection of the residuals suggests the
possible presence of emission features in the Fe K band.  In Fig.~2 we
show the data to model ratio in the Fe K band for the simple model
above.  The vertical line marks the energy of 6.4~keV (Fe K$\alpha$)
in the source rest--frame.

\subsection{Fe emission: simple models}

To account for the residuals in Fig.~2, we add to the model a Gaussian
emission line and find a significant improvement of $\Delta\chi^2=14$
for three more free parameters corresponding to a significance at the
$\sim$~99.9 per cent level according to the F--test for a final result
of $\chi^2=353$ for 395 dof. However, as pointed out by Protassov
    et al. (2002), the estimate of the statistical significance of emission
    lines through the F--test is not highly accurate. We then
    performed Monte Carlo simulations according to the method proposed
    in Porquet et al.  (2004) and our results confirm the significance
    of the Fe line detection at more than the 99.6 per cent confidence
    level.  The best--fitting parameters indicate a line energy of
$E=6.62\pm 0.18$~keV with a rest--frame equivalent width EW$=90 \pm
35$~eV with respect to the absorbed power law continuum. The line
appears to be barely resolved and we measure a width $\sigma=220 \pm
170$~eV.

However, Fig.~2 suggests the presence of a more structured line
profile and the possible presence of two emission lines. The brightest
is close to 6.4~keV, while the fainter is seen at higher energy. As a
first phenomenological model, we consider a double--Gaussian
parametrisation of the emission features. We obtain a slightly better
fit than with a single Gaussian ($\chi^2=348$ for 392 dof). One line
is found at $E=6.46^{+0.09}_{-0.05}$~keV and has an equivalent width
of 60~eV in the rest--frame. The second line is at
$E=6.86^{+0.11}_{-0.23}$~keV and is fainter. Both lines are unresolved
with an upper limit on their width of 200~eV and 300~eV respectively.
The energy of the lines indicates the presence of a slightly ionized
reflection component producing a $\sim$~6.4~keV neutral K$\alpha$ line
and a fainter He--like one.

\subsubsection{A relativistic Fe line}

A second parametrisation is suggested by previous X--ray observations
of 3C~109.  An Fe emission line was first detected in the X--ray
spectrum of 3C~109 with {\it ASCA} and result are reported by Allen et
al (1997). When fitted with a single Gaussian emission line, the {\it
  XMM--Newton} best--fitting parameters (see above) are different but
consistent within the 90 per cent errors with the {\it ASCA } results.
In particular a much larger width ($\sigma=650^{+810}_{-360}$~eV) and
equivalent width ($300^{+600}_{-200}$~eV) were suggested by the {\it
  ASCA} spectrum.  The Fe line detected by {\it ASCA} was broad and
consistent with being emitted from the accretion disc. The Fe line
emitting region was however poorly constrained and the inner disc
radius was allowed to take any value from the innermost stable
circular orbit (ISCO) around a non--rotating black hole up to
$140~r_g$ ($r_g=GM/c^2$: the ISCO for a non--rotating black hole is
$6~r_g$).

To directly compare the {\it XMM--Newton} data with the {\it ASCA}
results on the Fe line, we replace the Gaussian emission line with a
{\small DISKLINE} model (Fabian et al 1989) as in Allen et al (1997).
We fix the line energy to 6.4~keV and the emissivity to the energy
dissipation profile of a standard accretion disc ($r^{-3}$).  The
inner disc radius and observer inclination are free to vary while the
outer disc radius is fixed to $10^3~r_g$. We obtain a good fit with
the same statistical quality as the one with a Gaussian line (model A
in Table~1). The inner disc radius can take any value between the ISCO
around a non--rotating black hole and $27~r_g$, significantly
improving the {\it ASCA} $140~r_g$ constraint, while the observer
inclination is $i=43^\circ \pm 12^\circ$.  If the line energy is
instead fixed to be 6.7~keV, corresponding to He--like Fe emission,
only a lower limit on the inclination can be obtained ($i > 34^\circ$)
and the inner disc radius is completely unconstrained ($6~r_g <
r_{\rm{in}} < 855~r_g$). Results are thus not conclusive on the
presence of a relativistic Fe line in 3C~109. The data are consistent
with a relativistic Fe line from a disc extending all the way down to
the ISCO, but a truncated disc missing the innermost $20$--$30~r_g$
gravitational radii cannot be excluded.

The Fe line EW is smaller than in the {\it ASCA} data (see
Table~1, model A), but the line intensity is consistent with being
constant within the errors.  By considering the neutral (6.4~keV) Fe
line model, we measure an intensity of $(1.3\pm 0.6)\times
10^{-5}$~ph~cm$^{-2}$~s$^{-1}$ to be compared with the {\it ASCA}
result of $(2.1^{+4.3}_{-1.3} \times 10^{-5}$~ph~cm$^{-2}$~s$^{-1}$.

\begin{figure}
\begin{center}
 \includegraphics[width=0.32\textwidth,height=0.44\textwidth,angle=-90]{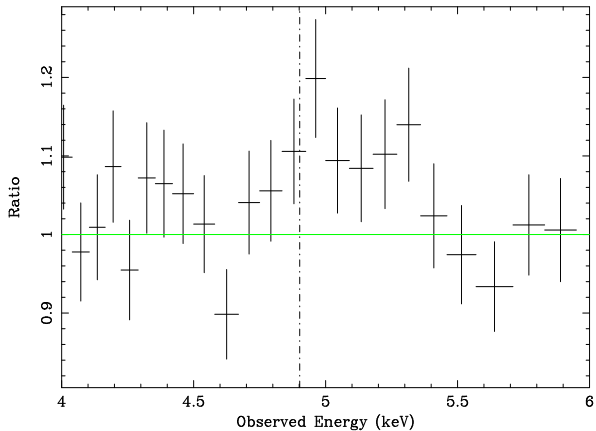}
\end{center}
\caption{Data to model ratio in the Fe K band. The model is a simple
  absorber power law fitted in the 2--10~keV band ignoring the
  Fe band. Data have been rebinned for visual clarity. The vertical
  line is at 6.4~keV in the source rest--frame.}
\end{figure}

\begin{figure}
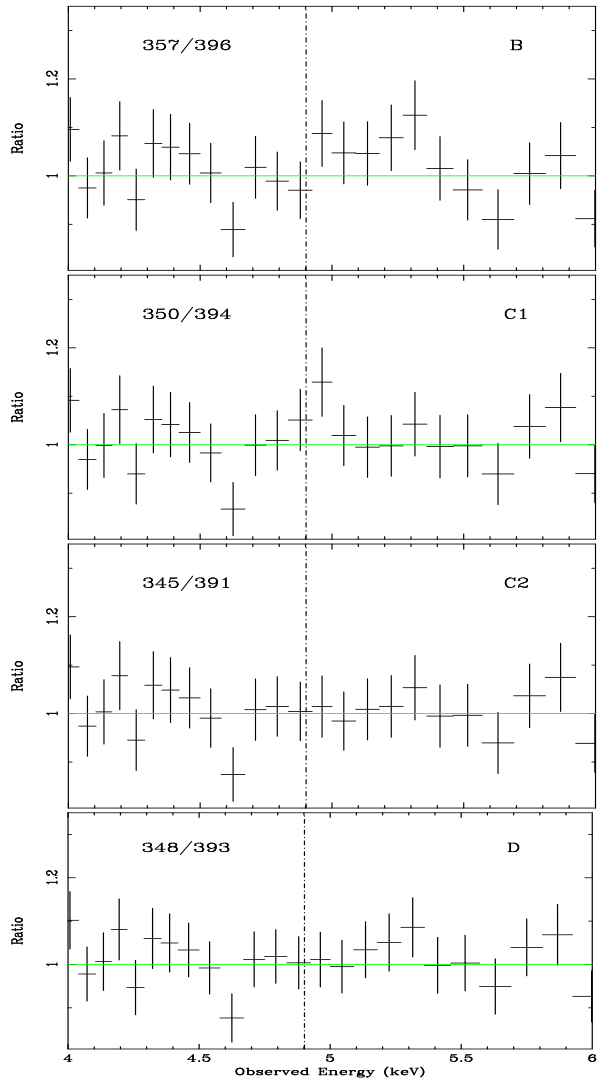

\begin{center}
 \includegraphics[width=0.2\textwidth,height=0.44\textwidth,angle=-90]{RatioB.cps}
 \includegraphics[width=0.2\textwidth,height=0.44\textwidth,angle=-90]{RatioC1.cps}
 \includegraphics[width=0.2\textwidth,height=0.44\textwidth,angle=-90]{RatioC2.cps}
 \includegraphics[width=0.2\textwidth,height=0.44\textwidth,angle=-90]{RatioD.cps}
\end{center}
\caption{We show the data to model ratio in the Fe K band for the
  different models used to describe the 2--10~keV spectrum of 3C~109
  (see text and Table~1 for details). From top to bottom models B, C1,
  C2, and D are shown together with the resulting statistics
  ($\chi^2$/dof). The vertical line is at 6.4~keV in the source
  rest--frame.}
\end{figure}


\subsection{Including the X--ray reflection continuum}

If the Fe line really originates in optically thick matter such as the
accretion disc (or the torus if narrow), the line is associated with a
reflection continuum that has to be included in the spectral model for
self--consistency. The X--ray reflection continuum introduces
curvature (and Fe edge) in the Fe band and could affect the line
parameters.  Moreover, when the reflection spectrum is ionized,
Comptonization induces a broadening of the line which has to be taken
into account.  We then apply to the hard spectrum a self--consistent
ionized reflection model in which the emission lines are computed
together with the X--ray reflection continuum (Ross \& Fabian 2005).
The model parameters are the Fe abundance and source redshift (that we
fix to the solar value and to the source redshift respectively), the
ionization parameter of the reflector, the photon index of an incident
power law continuum (which has a high--energy cutoff at 100~keV) and
the normalisation.

\subsubsection{Reflection from distant matter}

As a first step, we assume that the reflection spectrum and Fe
features come from distant material so that the only Fe line
broadening can be due to Comptonization if the reflector is highly
ionized. We obtain a good fit to the 2--10~keV pn spectrum (model B in
Table~1). The ionization parameter is constrained to be smaller than
$244$~erg~cm~s$^{-1}$. The best--fitting reflection spectrum comprises
both a neutral 6.4~keV line and a fainter He--like one at 6.7~keV. The
data to model ratio in the Fe band is shown in the top panel of
Fig.~3. Residuals are left above 6.4~keV (rest--frame, vertical line)
and we tried to account for them by adding a second more ionized
reflector, but we did not find any improvement in the statistics. We
measure a reflection fraction of $0.8\pm 0.2$, i.e. the reflector
intercepts about half (or only slightly less) of the X--ray continuum
in good agreement also with the measured Fe line equivalent width (see
model A in Table~1). 

\subsubsection{Accretion disc reflection}

We then consider the same reflection model as above, but we apply a
relativistic blurring kernel (obtained from the {\small DISKLINE}
model) inducing the relativistic effects arising if emission comes
from an accretion disc around a non--rotating black hole (model C1 in
Table~1). As for model A, we fix the emissivity to its standard
profile $r^{-3}$. We obtain an improvement of $\Delta\chi^2=7$ with
respect to the unblurred reflection model B for two more free
parameters which means that the relativistic blurring is significant
at the 98 per cent level. The inner disc radius is constrained to be
smaller than $26~r_g$, while the observer inclination is measured to
be $i = 40^\circ \pm 8^\circ$, confirming the results obtained with
the simpler power law plus {\small DISKLINE} model A when the line was
assumed to be neutral. As for the ionization state, we measure a low
ionization parameter $\xi < 30$~erg~cm~s$^{-1}$ which again points
towards a neutral (or slightly ionized) accretion disc. We also tried
to artificially increase the ionization of the reflector to reproduce
the results obtained with model A when the line energy was fixed to
6.7~keV but we could not find any solution. Once again, the
best--fitting solution points toward a significant contribution of the
reflection component (with a reflection fraction of the order of unity)
suggesting a typical Seyfert~1--like X--ray continuum.

\begin{figure}
\begin{center}
 \includegraphics[width=0.32\textwidth,height=0.44\textwidth,angle=-90]{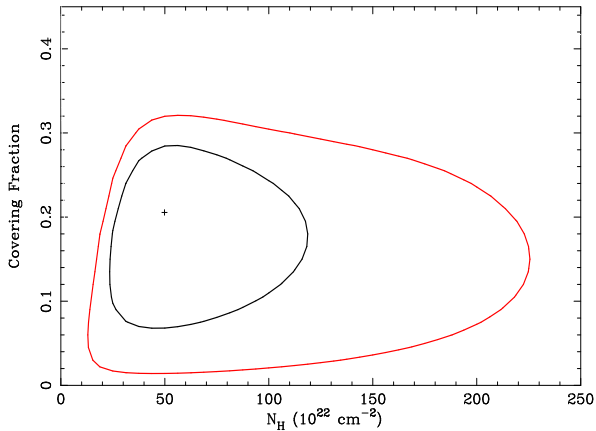}
\end{center}
\caption{Confidence level contours (68 and 90 per cent) for the
  covering fraction and the column density of the partial covering
  model to the 2--10~keV spectrum (model D).}
\label{bband}
\end{figure}

\subsubsection{Composite reflection (disc plus distant matter)}

Visual inspection of the residuals shows that the 6.4~keV peak is not
well modelled (second panel of Fig.~3). We then add to model C1 a
narrow Gaussian emission line representing the narrow component of the
Fe K line (model C2 in Table~1). A narrow Fe K line is in fact
ubiquitous in the X--ray spectra of Active galaxies and is thought to
be associated with some distant reflector such as the torus of
unification schemes (e.g.  Yaqoob \& Padmanabhan 2004). The inclusion
of the narrow emission line marginally improves the fit by $\chi^2=5$
for three more free parameters. The line parameters are reported in
Table~1 and the data to model ratio is shown in the third panel of
Fig.~3. The blurring reflector parameters (inner disc radius and
  inclination) are fitted to the data being however unaffected and
indistinguishable from results obtained with model C1. If the narrow
line is replaced by an unblurred reflection spectrum no further
improvement is obtained.

\subsection{Partial covering alternative to the relativistic Fe line}

The 2--10~keV spectrum is best fitted (though marginally) by model C2
comprising an almost neutral reflection spectrum from the accretion
disc plus a component from distant matter producing the narrow Fe line
at 6.4~keV. As mentioned, if relativistic blurring is removed the
statistical quality of the fit is only slightly worse and we are thus
motivated to find alternative solutions. Here we briefly explore one
further alternative to the relativistic line.
\begin{table*}
\begin{center}
  \caption{Summary of the best--fit parameters and 90 per cent errors
    ($\Delta\chi^2=2.71$ for one parameter) for the most relevant
    spectral models used to describe the X--ray spectrum of 3C~109.
    Only the EPIC--pn data are used in the 2--10~keV band, while the
    MOS 1 and MOS 2 data are included in the broadband (0.4--10~keV)
    analysis. The parameters refer to the 2--10~keV analysis but
    consistent results (within the errors) are obtained when the
    broadband spectrum is considered except for the photon index of
    model A that steepens to $\Gamma\simeq 1.75$ in the 0.4--10~keV analysis.}
\begin{tabular}{cccccccccc}
\hline
\multicolumn{1}{c}{MODEL} &
\multicolumn{1}{c}{} &
\multicolumn{6}{c}{PARAMETERS} &
\multicolumn{1}{c}{2--10~keV} &
\multicolumn{1}{c}{0.4--10~keV} \\
\multicolumn{1}{c}{} &
\multicolumn{1}{c}{} &
\multicolumn{6}{c}{} &
\multicolumn{1}{c}{$\chi^2$/dof} &
\multicolumn{1}{c}{$\chi^2$/dof} \\
\hline
A && $\Gamma$ & E$_{\rm{Fe}}$ &  r$_{\rm{in}}$ & i & EW$_{\rm{Fe}}$
&--& & \\
{\small POW + DISKLINE} && $1.68\pm 0.07$ & $6.4^f$ & $<27~r_g$ &
$43^\circ \pm 12^\circ$ & $170\pm 70$ &--& 353/395 & 1322/1364\\ \\
 && $1.68\pm 0.07$ & $6.7^f$ & $ 380^{+475}_{-374}~r_g$ & $> 34^\circ$
 & $105\pm 40$ &--& 351/395 & 1319/1364\\
\\
B && $\Gamma$ & --  & --  & --  &  $\xi$ &R & &\\
{\small POW + REF} && $1.75^{+0.06}_{-0.08}$ & -- &-- & -- & $< 244$
&$0.8\pm 0.2$& 357/396&  1317/1365\\
\\
C1 && $\Gamma$ & --&
r$_{\rm{in}}$ & i & $\xi$ &R&& \\
{\small POW + DISCREF} && $1.87\pm 0.10$ & -- & $< 26~r_g$ &
$40^\circ \pm 8^\circ$ & $< 30$ &$1.1\pm 0.2$& 350/394 & 1297/1363\\
\\
C2 && $\Gamma$ &  E$_{\rm{Fe}}$&
$\sigma$ & EW$_{\rm{Fe}}$ & $\xi$ &R& &\\
{\small POW + DISCREF + GAUSS} && $1.87\pm 0.10$ & $6.45^{+0.09}_{-0.06}$ & $<340$ &
$45 \pm 15$ & $<45$ &$1.1\pm 0.2$& 345/391 & 1291/1360\\
\\
D && $\Gamma$ & E$_{\rm{Fe}}$ &
$\sigma$ & EW$_{\rm{Fe}}$ &  N$_{\rm{H}}$ &f$_{\rm{c}}$& & \\
{\small PC ( POW + GAUSS ) } && $1.86\pm 0.08$ & $6.47^{+0.14}_{-0.05}$
& $<325$ & $50^{+60}_{-30}$ & $500^{+1800}_{-390}$ &$<32\%$& 348/393 &
1292/1362\\
\hline
\end{tabular}
{\vspace{0.2cm}}
\parbox {7in} {All fits include photoelectric absorption (the {\small
    WABS} model). The resulting column density is not reported here.
  It is always of the order of $3$--$5\times 10^{21}$~cm$^{-2}$ and a
  detailed analysis of the absorption in 3C~109 is given below (see
  Section~5 and Table~2). In models A and D the Fe line energy
  E$_{\rm{Fe}}$ is given in keV in the source rest--frame, while the
  line width and equivalent width (EW) are given in eV (the EW is also
  corrected for the source redshift). In model B, C1, and C2 {\small
    REF} refers to the unblurred reflection model, and {\small
    DISCREF} to the relativistically blurred one. The ionization
  parameter of the reflection models ($\xi$) is given in units of
  erg~cm~s$^{-1}$ and R is the reflection fraction (R=1 means that the
  reflector intercepts about half of the X--ray continuum). Given that
  our reflection model does not contain the reflection fraction as a
  free parameter, its value is estimated by comparison of the
  best--fitting reflection model with the {\tt PEXRAV} model (neutral
  reflection), a procedure that is reasonably accurate here given that
  the ionization state of the reflector is low. In model C2 the
  parameters $r_{\rm{in}}$ and $i$ are the same as in model C1 and are
  not reported. In model D, the column density of the partial coverer
  is given in units of $10^{21}$~cm$^{-2}$.}
\label{diskline}
\end{center}
\end{table*}

We consider the possibility that absorption by a large column of gas
partially covering the source is responsible for the hard
spectral shape, as tentatively observed in other radio--loud sources
(e.g. Lewis et al. 2005). We also include a Gaussian emission line to
account for possible Fe emission from the absorber itself (or other
distant material). We obtain a good fit with a partial covering model
(model D) which is statistically indistinguishable from our
best--fitting model C2 (see Table~1).  The data to model ratio for
model D are shown in the bottom panel of Fig.~3.  The partial covering
parameters are however not very well constrained. The covering
fraction is only an upper limit of 32 per cent and the column density
can take any value from $10^{23}$~cm$^{-2}$ to $2.3\times
10^{24}$~cm$^{-2}$ (2$\sigma$ errors). The 68 and 90 per cent
confidence level contours for covering fraction and column density are
shown in Fig.~4. We also replaced the neutral partial coverer with an
ionized one, but the requirement is for neutral matter with a 90 per
cent upper limit on the ionization parameter of $\xi < 0.5$~erg~cm~s$^{-1}$.

\begin{figure}
\begin{center}
 \includegraphics[width=0.32\textwidth,height=0.44\textwidth,angle=-90]{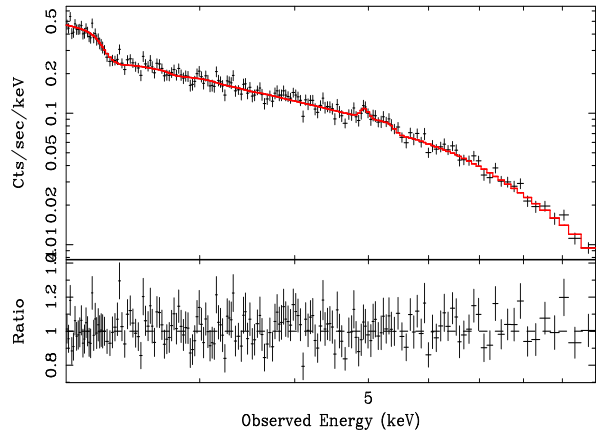}
\end{center}
\caption{We show the 2--10~keV pn spectrum and data to
  model ratio for model C2 (see text and Table~1).}
\label{hard}
\end{figure}

\subsection{Conclusions and estimate of the Eddington ratio}

We conclude that the spectral shape in the hard band is well
reproduced by either {\emph{i)}} an almost neutral X--ray reflection
spectrum from the accretion disc with a likely contribution from a
distant reflector (model C2) or by {\emph{ii)}} intervening absorption
by a large column of gas ($> 10^{23}$~cm$^{-2}$) covering about 20 per
cent of the X--ray continuum source plus a narrow Fe K$\alpha$ line
(model D). Both interpretations are allowed by the data and cannot be
distinguished on a statistical basis. In Fig.~5 we show the 2--10~keV
spectrum and data to model ratio when the spectrum is fitted with
model C2.

By choosing C2 as best--fit model , we measure an observed 2--10~keV
flux of $6.8\times 10^{-12}$~erg~cm$^{-2}$~s$^{-1}$. Correcting for
absorption, we have a 2--10~keV flux of $7\times
10^{-12}$~erg~cm$^{-2}$~s$^{-1}$, corresponding to a 2--10~keV
luminosity of $2\times 10^{45}$~erg~s$^{-1}$ in the source
rest--frame. If partial covering is instead the right description of
the hard spectrum, the intrinsic absorption corrected luminosity is
$\simeq 3\times 10^{45}$~erg~s$^{-1}$, depending on the assumed column
density ($H_0=70$~km~s$^{-1}$~Mpc$^{-1}$,
  $\Omega_\Lambda=0.73$, $\Omega_0=1$). 

Given the recent mass estimate for the central black hole
($M_{\rm{BH}} \simeq 2\times 10^8~M_\odot$, McLure et al 2006), and by
assuming a 2--10~keV to bolometric luminosity conversion\footnote{The
  adopted bolometric correction of 30 for the 2--10~keV luminosity is
  a conservative one and is likely to underestimate the actual
  bolometric luminosity of 3C~109 by a factor 2--3, see e.g.  Elvis,
  Risaliti \& Zamorani 2002. Therefore the Eddington ratio discussed
  below has to be intended as a lower limit.} of a factor 30 (Fabian
\& Iwasawa 1999 from the compilation of spectral energy distributions
of Elvis et al.  1994), it appears that 3C~109 is characterised by a
super--Eddington luminosity with $L_{\rm{bol}}/L_{\rm{Edd}} \simeq
2.5$. The black hole mass estimate is crucial for this result: McLure
et al. (2005) used the H$\beta$ version of the virial--mass estimator
(McLure \& Dunlop 2004) which provides black hole mass estimates with
a known scatter of $\sim$~0.4~dex, meaning that the black hole mass of
3C~109 could be as large as $\sim 5\times 10^8~M_\odot$. However, even
by adopting this upper limit, the requirement is still for a source
radiating at its Eddington luminosity. 

The above discussion assumes that the X--ray luminosity is radiated
isotropically and, given the FR~II nature of 3C~109, this is far from
obvious. It is indeed possible that part of the X--ray core emission
is due to a beamed inner jet component, which would reduce the
$L_{\rm{bol}}/L_{\rm{Edd}}$ ratio. The jet component is generally
associated with a flatter (1.5--1.7) X--ray slope than measured here
(1.86). However, it should be stressed that some of the knots in the
jets of FR~II radio galaxies can have much steeper photon--indexes
than generally seen ($\Gamma\sim 2$ though with large error bars, see e.g.
Sambruna et al. 2006). Thus the steep photon index we measure cannot
be taken as a secure indication that beamed emission is negligible. On
the other hand, if reflection rather than partial covering is the
rigth description of the hard X--ray data, the measured reflection
fraction (of the order of unity) implies that the X--ray continuum is
not strongly beamed away from the reflector, thus suggesting little
inner jet contamination. We conclude that, though it is difficult to
exclude that the inferred bolometric luminosity is over--estimated
because of the presence of a beamed component, the possibility that
3c~109 is radiating a substantial fraction (close to unity) of its
Eddington luminosity does remain plausible, which would also be
consistent with the tentative detection of the broad Fe line from the
accretion disc.

\section{Broadband spectral analysis}

We then consider the broadband spectrum of 3C~109.  We use the full
0.4--10~keV band and consider joint fits to all EPIC cameras (pn and
the two MOS). The data below 0.4~keV are ignored because of some
residual inter--calibration problems at the softest energies. The
extension of the 2--10~keV spectral models to the soft band is
satisfactory (even before fitting the data) for all models. In the
last column of Table~1 we report the results of the fits to the
broadband spectrum with the same models used for the hard band. The
parameters reported in Table~1 refer to the 2--10~keV fits but are
basically unaffected by the inclusion of the low energy data except
for the photon index of models A which steepens to $\Gamma\simeq 1.75$.

Once again the data tend to prefer models C2 and D as the
best--fitting solutions which are indistinguishable from a statistical
point of view. We notice that models C1, C2, and D have a steeper
spectral index than models A and B. This is the main reasons for the
better fits provided by models C1, C2 and D. In fact if only the soft
data below 2.5~keV are considered (where the reflection/partial
covering contribution is negligible) a fit with an absorbed power law
gives $\Gamma\simeq 1.85$ confirming that a steep photon index is
preferred by the data as in models C1, C2, and D and suggesting that
models A and B have to be considered as less likely as also indicated
by the fitting statistics.

\subsection{Long--term X--ray variability}

The broadband spectral analysis is in excellent agreement with the
{\it ASCA} data presented in Allen et al. (1997) and enable us to
compare the observed flux with previous X--ray observations. We
consider the 1--2~keV flux only, since this energy band is common to
all previous observations. We measure a flux of $(14.5\pm0.3)\times
10^{-13}$ and, in Fig.~6, we show the historical X--ray flux light
curve in that same energy band. All historical data are re--collected
from Allen et al (1997) with the exception of a second 1998 {\it ASCA}
observation (unpublished) and of the 2003 Chandra observation
(Hardcastle, Evans \& Croston  2006) that we include here for the first time. Long--term
X--ray variability is clearly present with the largest amplitude
variation occurring in a 4 years interval between the 1991 {\it ROSAT}
and the 1995 {\it ASCA} observations (about a factor 2).  Variability
with the similar amplitude and timescale has also been reported in the
near--infrared (Rudy et al.  1984) and variation of the order of 50
per cent are seen in the J, H, and K bands as well (Elvis et al.
1984). 

\begin{figure}
\begin{center}
 \includegraphics[width=0.32\textwidth,height=0.44\textwidth,angle=-90]{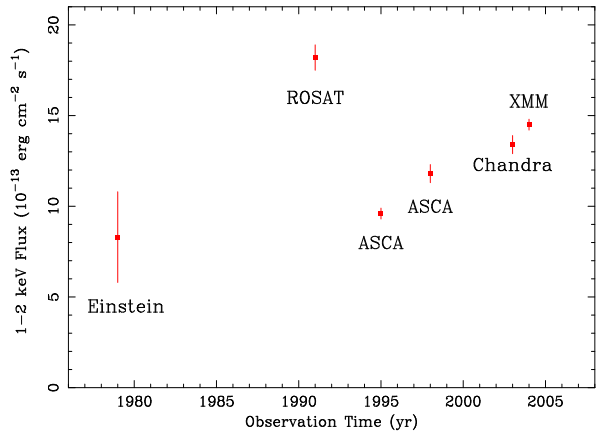}
\end{center}
\caption{The historical X--ray flux of 3C~109 in the 1--2~keV
  band.}
\label{GammaNH}
\end{figure}

\section{Excess absorption}

When model D is extended to the 0.4--10~keV band, we measure (besides
the partial covering column density) a column density of N$_{\rm{H}} =
4.64^{+0.09}_{-0.21}\times 10^{21}$~cm$^{-2}$ which is derived by
assuming that the absorber lies at zero redshift. An identical result
is obtained by using model C2 instead. 

The value of the Galactic column density in the line of sight towards
3C~109 is somewhat ambiguous. The nominal value obtained through H I
measurements is $1.57\times 10^{21}$~cm$^{-2}$ (Burstein \& Heiles
1982). However, COBE/IRAS maps indicate a Galactic column as high as
$3.31 \times 10^{21}$~cm$^{-2}$ (Schlegel, Finkbeiner \& Davis 1998),
in good agreement with X--ray studies of the nearby cluster Abell~478
suggesting a column of $\sim 2.5-3\times 10^{21}$~cm$^{-2}$ (Sun et
al. 2003; Pointecouteau 2004; Sanderson, Finoguenov \& Mohr 2005). The
same result is obtained by averaging the COBE/IRAS reddening over a
1~deg$^2$ region centred on 3C~109 and, since this result is based on
direct observations of the dust, we adopt this value as already done
by Rudy, Puetter \& Mazuk (1999) in their infrared study in which the
{\emph{total}} column density towards 3C~109 was estimated to be $\sim
7.7\times 10^{21}$~cm$^{-2}$.

The column density we measure in the broadband fits
($4.64^{+0.09}_{-0.21}\times 10^{21}$~cm$^{-2}$) strongly suggests
that excess absorption is present, confirming previous analysis of
{\it ROSAT PSPC} and {\it ASCA} data (Allen \& Fabian 1992; Allen et
al 1997). Assuming, as we did so far, that the absorber lies at zero
redshift, a column density of $(5.30\pm 0.42)\times 10^{21}$~cm$^{-2}$
was inferred from previous {\it ASCA} spectral analysis, while {\it
  ROSAT} data indicates a total column density of $4.2^{+1.9}_{-1.6}
\times 10^{21}$~cm$^{-2}$. The {\it XMM--Newton} data confirm the
presence of excess absorption in the line of sight with a (zero
redshift) column density somewhat lower than the {\it ASCA} result and
consistent with the {\it ROSAT} spectrum.

Since we are here interested in determining the properties of an
absorber with column density of a few times $10^{21}$~cm$^{-2}$ and
given that such a column does not affect significantly the spectrum
above 3~keV, in the following we consider the 0.4--2.5~keV band only.
In this way, we are not affected by the spectral complexities in the
hard band (partial covering and/or reflection) and we are certain of
measuring the intrinsic spectral slope, whose value clearly affects
the absorption properties. In Fig.~7 we show the confidence level
contours for the power law photon index and total column density
(model I in Table~2). The contours confirm that the total column
density seen in the X--ray spectrum of 3C~109 is well in excess of all
estimates for the Galactic value.
\begin{figure}
\begin{center}
 \includegraphics[width=0.32\textwidth,height=0.44\textwidth,angle=-90]{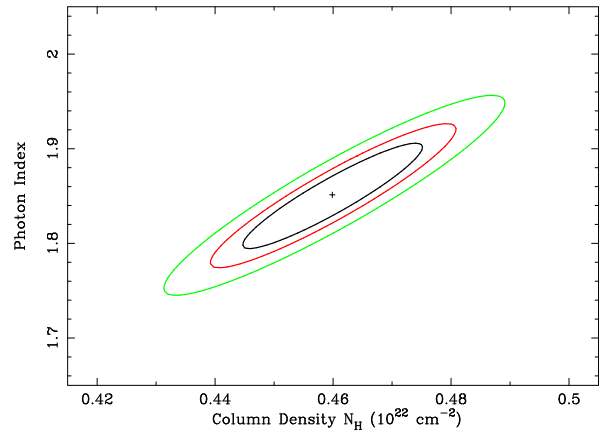}
\end{center}
\caption{Confidence contours for photon index and total column density
  from the absorbed power law model fit to the pn and MOS data in the
  0.4--2.5~keV band (model I in Table~2). Contours represent the 68,
  90, and 99 per cent confidence levels.}
\label{GammaNH}
\end{figure}

\subsection{Excess absorption local to 3C~109 ?}

\begin{table*}
\begin{center}
  \caption{Summary of the best--fitting parameters and 90 per cent
    errors for the different absorption models used to describe the
    0.4--2.5~keV spectrum of 3C~109. Data from the EPIC pn and MOS
    cameras are fitted jointly. }
\begin{tabular}{cccccccccc}
\hline
\multicolumn{1}{c}{MODEL} &
\multicolumn{6}{c}{PARAMETERS} &
\multicolumn{1}{c}{$\chi^2$/dof} \\
\hline
I && $\Gamma$ & N$_{\rm H}$ & -- & -- & -- & \\
{\small WABS * POW } && $1.85^{+0.04}_{-0.06}$ & $4.60\pm 0.17$&
-- & -- &--& 613/632\\
\\
II && $\Gamma$ & N$_{\rm H}$ & N$_{\rm H} (z)$  & $z$ & -- & \\
{\small WABS * ZWABS * POW } && $1.84^{+0.06}_{-0.04}$ &
$3.31^f$&$1.20^{+0.25}_{-0.10}$ & $<0.02$ &--& 612/631\\ \\
&& $1.84 \pm 0.07$ & $4.57^{+0.10}_{-0.58}$&$<0.9$ & $0.3056^f$ &--&
613/631
\\
\\
III && $\Gamma$ & N$_{\rm H}$ & N$_{\rm H} (z)$ & $z$ & $\xi_{\rm{abs}}$ & \\
{\small WABS * ABSORI * POW } && $1.85^{+0.07}_{-0.06}$ & $3.31^f$&
$3.00^{+0.67}_{-0.40}$ & $> 0.28$ &$0.55^{+0.25}_{-0.40}$& 600/630\\
\\
&& $1.86^{+0.07}_{-0.05}$ & $3.50\pm 0.55$&
$2.80\pm 1.25$ & $0.3056^f$ &$0.70^{+0.55}_{-0.35}$& 599/630
\\
\\
IV && $\Gamma$ & N$_{\rm H}$ & N$_{\rm H} (z)$ & $z$ & $\xi_{\rm{abs}}$ & \\
{\small WABS * XSTAR * POW } && $1.91^{+0.07}_{-0.06}$ & $3.36^{+0.41}_{-0.62}$&
$4.00^{+2.30}_{-1.80}$ & $0.3056^f$ &$0.39^{+0.43}_{-0.17}$& 588/630\\
\hline
\end{tabular}
{\vspace{0.2cm}}
\parbox {7in} {All column densities are given in units of
  $10^{21}$~cm$^{-2}$. The ionization parameter ($\xi$) is given in
  units of erg~cm~s$^{-1}$.}
\end{center}
\end{table*}

Part of the excess absorption could be due to material surrounding the
active nucleus and therefore at the redshift of 3C~109. We first add a
second neutral absorber with redshift free to vary between zero and
the source redshift (0.3065). Given the uncertainties in the Galactic
absorption, we make a conservative choice and we assume the highest
estimated value for the Galactic column density in the line of sight
($3.31\times 10^{21}$~cm$^{-2}$). We obtain a good fit but the
redshift of the second absorber is only a 90 per cent upper limit
($z<0.02$) ruling out the presence of a neutral absorber at the redshift
of the source. The same result is obtained if the Galactic column
density is free to vary and the second absorption component is imposed
to arise at z=0.3056 (see Table~2, model II). In this case, it is the
column density of the absorber at z=0.3056 that is only an upper limit
($< 9\times 10^{20}$~cm$^{-2}$).

\subsection{A ``tepid absorber'' solution}

The presence of excess absorption with column densities of the order
of few times $10^{21}$~cm$^{-2}$ (or more) is often detected in the
X--ray spectra of Active Galactic Nuclei. Given the luminosities of
the central engine, the absorbing gas is almost always ionized and
gives rise to typical signatures in the soft X--ray band.  The soft
X--ray spectrum of 3C~109 does not present any clear indication that a
warm absorber is present.  However, if the ionization parameter of the
absorbing gas is low (but different from zero) subtle differences are
expected with a neutral absorber without strong signatures in the soft
band. 

To test the hypothesis that the excess absorption in 3C~109 is
slightly ionized, we use the simple {\small ABSORI} model in {\small
  XSPEC} (Done et al. 1992). As in model II, we first fix the column
density of the (neutral) zero--redshift absorber to the Galactic
estimate and let the ionized absorber redshift be a free parameter
between 0 and 0.3056 (model III in Table~2). We find a significant
improvement with $\Delta\chi^2 = 12$ for one more free parameters with
respect to model II (a significance of more than 99.95 per cent). The
absorber is only slightly ionized with
$\xi=0.55^{+0.25}_{-0.40}$~erg~cm~s$^{-1}$ and the 90 per cent lower
limit on the redshift of the ionized absorber is $z>0.28$ indicating
an {\emph{in situ}} origin. With the ionized solution, we measure a
column density of $3.00^{+0.67}_{-0.47} \times 10^{21}$~cm$^{-2}$ in
excess of the Galactic. We then remove the degeneracy between
ionization parameter and redshift by fixing the redshift of the
ionized absorber to z=0.3056 and let the Galactic column density free
to vary.  We find a similar fit as the one above with $N_H^{\rm{Gal}}
= (3.50\pm 0.55) \times 10^{21}$~cm$^{-2}$ and $N_H^{\rm{z}} =
(2.80\pm 1.25) \times 10^{21}$~cm$^{-2}$, while the ionized absorber
has a ionization parameter of
$\xi=0.70^{+0.55}_{-0.35}$~erg~cm~s$^{-1}$.

As a final
test, we consider a more refined absorption model using the {\small
  XSTAR} code (Kallman \& Bautista 2001). We find an excellent fit to
the 0.4--2.5~keV data with a significant improvement on any previous
spectral model (see Table~2, model IV). Our best--fitting model is
obtained by fixing the ionized absorber redshift only, while the
Galactic and {\emph{in situ}} column densities and the absorber
ionization parameter are free to vary. We measure a zero--redshift,
neutral column of $N_H^{\rm{Gal}} = 3.36^{+0.41}_{-0.62} \times
10^{21}$~cm$^{-2}$, while the ionized absorber has a column density of
$N_H^{\rm{z}} = 4.00^{+2.30}_{-1.80} \times 10^{21}$~cm$^{-2}$ with a
ionization parameter of $\xi=0.39^{+0.43}_{-0.17}$~erg~cm~s$^{-1}$.

In Fig.~8 we show the confidence contours for the ($\log$ of the)
ionization parameter and the column density of the ionized absorber
from our best--fitting model IV. The absorber is only very slightly
ionized, but the ionization parameter is well defined. Our best--fit
model (IV) thus comprises a neutral zero--redshift plus ionized
absorption at the redshift of 3C~109 for a total column density of
$\sim 7.4 \times 10^{21}$~cm$^{-2}$ which is in excellent agreement
with the total obscuration inferred through infrared studies (and
implying a column density of $\sim 7.7\times 10^{21}$~cm$^{-2}$).

\begin{figure}
\begin{center}
 \includegraphics[width=0.32\textwidth,height=0.44\textwidth,angle=-90]{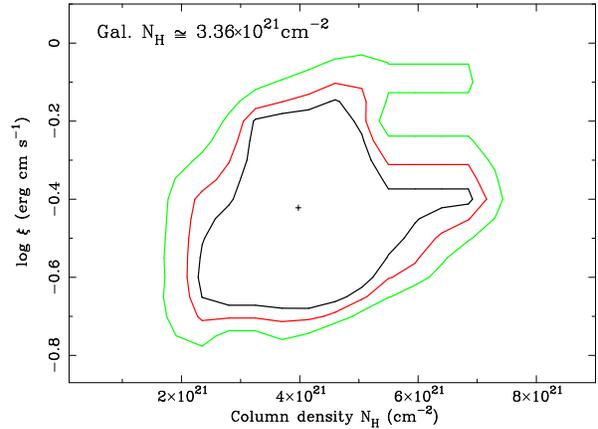}
\end{center}
\caption{We show confidence contours for the ionization parameter and
  column density of the ionized absorber modelled through {\small
    XSTAR} (model IV, Table~2). The additional neutral and
  zero--redshift absorber provides an additional column density of
  $3.36^{+0.41}_{-0.62} \times 10^{21}$~cm$^{-2}$ making the total
  column density of the order of $\sim 7.4 \times 10^{21}$~cm$^{-2}$,
  in excellent agreement with studies in the optical and infrared.}
\end{figure}

By applying our best--fit absorption model (IV) to the broadband model C2
(see Table~1) we obtain a final fit to the broadband 0.4--10~keV
spectrum of 3C~109 with $\chi^2=1272$ for 1358 dof. The same statistical
quality is obtained by considering model D instead. No significant
change is seen in the best--fit parameters reported in Table~1 and 2.

\section{Discussion}

We have presented results from a new $\sim$40~ks {\it XMM--Newton}
observation of the BLRG/QSO 3C~109 (z=0.3056), one of the most X--ray
luminous sources within redshift 0.5. The intrinsic X--ray luminosity
of the source is found to be $\sim 2$--$3\times 10^{45}$~erg~s$^{-1}$
(depending on the adopted model) and compares well with previous
X--ray estimates, though the source is shown to be variable by a
factor 2 at least on a few years timescale. The X--ray continuum
    is well represented by a relatively steep power law with
    $\Gamma\simeq 1.86$. Such a steep photon index is generally taken
    as an indication that a beamed component (often characterised by a
    flatter 1.5--1.7 photon index) contributes only marginally to the
    X--ray spectrum. However, this is not always the case, and an
    inner jet component cannot be ruled out on the basis of a steep
    X--ray slope only. The beamed component contribution to the X--ray
    spectrum is particularly relevant because by combining the high
    X--ray luminosity with the black hole mass estimate ($\sim 2\times
    10^8~M_\odot$), 3C~109 appears to be characterised by
    $L_{\rm{bol}}/L_{\rm{Edd}} \simeq 2.5$ which can be significantly
    reduced only by assuming that the X--ray spectrum is strongly
    contaminated by a beamed inner jet component.

Fe emission is seen in the 6.3--7~keV rest--frame band. A narrow core
at 6.4~keV is detected unambiguously, while additional curvature in
the Fe K band can be modelled either in terms of reflection from the
accretion disc or by invoking a partial covering scenario. If the disc
reflection interpretation is correct, the disc surface is only
slightly ionized and the emitting region must extend down to at least
$\sim$~26 gravitational radii from the black hole. Such a solution
allows one to place the inner disc radius as close as the innermost
stable circular orbit around a Schwarzschild black hole (6
gravitational radii) but is also consistent with a truncated disc
scenario in which the innermost 20--30 gravitational radii are
missing, possibly replaced by thick and radiatively inefficient
matter. The observer inclination is well constrained ($40^\circ \pm
8^\circ$) and is consistent with 3C~109 being a BLRG/QSO borderline
object, as indicated also by optical, infrared and radio studies. We
measure a reflection fraction of the order of unity which is
completely consistent with an overall 100~eV equivalent width of the
Fe line (narrow core plus disc line). The detection of substantial
X--ray reflection from the accretion disc suggests that the X--ray
continuum is not significantly beamed away from the reflector and
supports the estimated $L_{\rm{bol}}/L_{\rm{Edd}} \simeq 2.5$
obtained by assuming isotropic X--ray emission. If instead partial
covering is the right description of the spectral curvature in the Fe
K band, about 20--30 per cent of the X--ray continuum source is
covered by a large column ($\sim 5\times 10^{23}$~cm$^{-2}$) of
neutral gas ($\xi < 0.5$~erg~cm~s$^{-1}$), and the intrinsic X--ray
luminosity still requires $L_{\rm{bol}}/L_{\rm{Edd}} > 1$ (but again
in the hypothesis of negligible inner jet contribution).

In the soft band, the X--ray continuum is clearly absorbed by a column
density in excess of the Galactic one, confirming previous findings
both in the X--rays and at optical and infrared wavelengths (Allen \&
Fabian 1992; Goodrich \& Cohen 1992; Allen et al. 1997; Rudy Puetter
\& Mazuk 1999). We obtain a total column density of the order of
$6$--$8 \times 10^{21}$~cm$^{-2}$ which has to be considered as the
sum of a zero--redshift neutral absorber and an {\emph{in situ}} one
at the redshift of 3C~109. The best--fitting solution to the pn and
MOS soft data strongly suggests that the {\emph{in situ}} absorber is
slightly ionized ($\xi \sim 0.4$~erg~cm~s$^{-1}$) with a non--zero
ionization parameter preferred at more than the 99.9 confidence level.
Our final result is that the zero--redshift neutral absorber
contributes with a column density of $\sim 3.4 \times
10^{21}$~cm$^{-2}$ to be added to a further $\sim 4.0 \times
10^{21}$~cm$^{-2}$ from the ionized absorber at the redshift of
3C~109. The total column density compares remarkably well with the
value of $\sim 7.7 \times 10^{21}$~cm$^{-2}$ inferred from optical and
infrared studies (Rudy, Puetter \& Mazuk 1999). Given the high
luminosity of 3C~109, the low ionization of the absorber places it far
from the central source (e.g. at the torus scale, suggesting that our
line--of--sight is grazing the torus). This is also consistent with
the observed large Balmer/Pashen decrement (Goodrich \& Cohen 1992;
Rudy Puetter \& Mazuk 1999) in the broad--line region (BLR) making it
likely that the {\emph{in situ}} absorber is further away than the BLR
with respect to the central engine. The obvious interpretation is that
our line of sight is grazing the molecular torus, in good agreement
with zeroth--order unification models (and with the inclination
inferred from the broad Fe line).

The excellent agreement between the X--ray and the optical/infrared
column density estimates implies that, if partial covering is indeed
the right description of the hard spectrum, the additional column of
neutral gas ($\sim 5\times 10^{23}$~cm$^{-2}$) cannot contribute to
the optical/infrared absorption and must i) be dust--free (i.e. it
lies within the dust sublimation radius), and ii) not obscure the
broad lines (i.e. it lies within the broad--line--region). This places
the partial coverer close to the X--ray source and within the
innermost $\simeq 0.1$~pc. By combining this location with the source
luminosity and the upper limit on the ionization parameter of the
absorber we estimate a density of the order of $n > 10^{11}$~cm$^{-3}$
for the gas responsible for the partial covering. Given the relatively
low covering fraction (20--30 per cent) and the large observed column
density ($\sim 5 \times 10^{23}$~cm$^{-2}$) the absorbing matter is
most likely distributed in compact clouds $< 5\times 10^{12}$~cm in
size. Given the black hole mass estimate, the cloud size is of the
order of $<0.17$ gravitational radii and to provide effective partial
covering, the clouds must have a size comparable to that the main
X--ray emitting region. However, according to standard accretion
models, X--rays mostly come from the innermost 50 gravitational radii
or so and thus, in order to provide a 20--30 per cent covering
fraction, the absorber must comprise hundreds of clouds in the line of
sight (therefore probably thousands of clouds if isotropically
distributed), which seems quite extreme a requirement.

We conclude that, even if not unambiguous from a spectral point of
view, the disc reflection (and broad Fe line) interpretation of the
hard spectrum of 3C~109 is less contrived than the partial covering
one and is also consistent with the high mass accretion rate implied
by the inferred $L_{\rm{bol}}/L_{\rm{Edd}}$. If partial covering is
ruled out, the contribution of the X--ray reflection spectrum to the
total flux implies a reflection fraction of the order of unity
suggesting the a putative inner jet component does not contribute
significantly in the X--ray band and further supporting the idea that
3C~109 is radiating close to its Eddington luminosity due to accretion
processes alone. The X--ray data do not rule out a truncated accretion
disc in which the innermost 20--30 gravitational radii are missing or
replaced by a radiatively inefficient flow, but this scenario is also
at odds with the high estimated Eddington ratio. If the innermost
regions of the accretion disc are instead present but highly ionized,
little Fe line is produced there, reconciling the hard spectral shape
with the huge X--ray luminosity and the high mass accretion rate
without the need to invoke a rather unusual partial covering scenario.

\section*{Acknowledgements}
Based on observations obtained with XMM-Newton, an ESA science mission
with instruments and contributions directly funded by ESA Member
States and NASA. GM thanks Ross McLure for useful discussions on the
black hole mass estimate in 3C~109 and Elena Belsole for discussions
on the extend X--ray emission. GM thanks the PPARC for support.  DRB
is supported by the University of Arizona Theoretical Astrophysics
Program Prize Postdoctoral Fellowship. This work was supported in part
by the U.S.  Department of Energy under contract number
DE-AC02-76SF00515. ACF thanks the Royal Society for support. RRR
thanks the College of the Holy Cross for support.

\end{document}